\documentstyle[aps,prb]{revtex}

\begin{document}
\title{Femtosecond data storage, processing and search using collective excitations
of a macroscopic quantum state.}
\author{D.Mihailovic, D.Dvorsek, V.V.Kabanov and J.Demsar}
\address{Jozef Stefan Institute, Jamova 39, 1000 Ljubljana, Slovenia}
\author{L.Forr\'{o}, H.Berger}
\address{Ecole Polytechnique Federale de Lausanne, CH-1015 Lausanne,\\
Switzerland}
\maketitle

\begin{abstract}
An ultrafast paralell data processor is described in which amplitude mode
excitations of a charge density wave (CDW) are used to encode data on the
surface of a 1-$T$ TaS$_{2}$ crystal. The data are written, manipulated and
read using parallel femtosecond laser pulse beams, and the operation of a
database search algorithm is demonstrated on a 2-element array.
\end{abstract}

\newpage Recently it was shown that the collective amplitude mode (AM)
oscillations of a charge-density wave state (CDW) can be detected in real
time in the quasi one-dimensional semiconductor K$_{0.3}$MoO$_{3}$ by
femtosecond time-resolved pump probe experiments\cite{Demsar}. In these
experiments, the AM is first excited by an ultrashort laser pump pulse which
acts as a $\delta -$function-like perturbation, and the resulting
reflectivity oscillations due to the modulation of the dielectric constant
by an AM oscillation are detected by suitably\ delayed probe pulses.

In the experiments presented in this letter, we demonstrate the feasibility
of using the collective AM excitation of a CDW in combination with
femtosecond-pulse laser beams to represent data written on the surface of a $%
1T$-TaS$_{2}$ crystal. We also proceed to perform basic data processing such
as storage, manipulation and readout operations. Because of the macroscopic
nature of the excitation, the operations can be performed at temperatures
approaching 200K with subpicosecond switching and readout times. Finally we
demonstrate a parallel search operation using a single pulse.

1{\it T-}TaS$_{2}$ is a layered nearly two-dimensional metal at room
temperature, which undergoes a transition to a CDW state driven by a Fermi
surface instability at $2k_{F}$ ($k_{F}$ is the Fermi wavevector) at
transition temperatures near $T_{c}\simeq 200$K. The macroscopic CDW state
is characterised by two collective excitations:\ the collective amplitude
(AM)\ and phason modes (PM), which are described by the solutions of the
quantum-mechanical Fr\"{o}hlich Hamiltonian\cite{Gruner}. Importantly from
the point of view of data processing, the AM decay time in this system is
quite long, $\tau _{p}$ $>10$ ps. Since this is almost two orders of
magnitude longer than the time necessary to excite the mode $\tau _{s}$ ($%
\sim 100$ fs), it gives us the opportunity to store data (in the form of
phase and amplitude of the AM) and also manipulate the AM database in
real-time with femtosecond laser pulses.

In what follows we first demonstrate the manipulation of the AM amplitude
and phase using different laser pulse sequences, an approach similar to
coherent control of coherent optical phonons performed on GaAs and GaAs/AlAs
superlattices \cite{Dekorsy,Dekorsy2}, Bi films \cite{Hase} and on
crystalline quartz \cite{Wefers}. We demonstrate switching of the AM on and
off at will, and show that we can set and detect the phase and amplitude of
the AM to a significant degree of accuracy. In effect, we show that the
collective amplitude mode excitation can be used as a switchable THz optical
modulator whose amplitude and phase can be controlled by laser pulses. In
the second step we demonstrate the operation of a search algorithm with a
single-query parallel search on multiple cells defined by multiple laser
spots on the sample.

A schematic experimental setup is shown in Figure 1. A stream of $80$ fs
pulses from a Ti:Sapphire laser at a wavelength of 800 nm is split into a
number of beams by Michaelson interferometers in which the time-delay (TD)$%
\;\Delta t$ between the pulses in different arms can be controlled to
sub-femtosecond precision. In the standard experimental configuration, one
or more pulse trains are incident on a spot on the surface of the crystal (a
cell), with which the amplitude and phase of the AM is written or changed,
while a strongly attenuated readout pulse beam is used to determine the
state of the cells by measurement of the reflectivity modulations as a
function of time. The experimental details have been described elsewhere\cite
{Demsar,Falques}.

The reflectivity modulation $\psi =\Delta R/R$ at the surface of the crystal
caused by a laser pump pulse at $t=0$ is shown in Figure 2a) (trace {\sf A}%
). Typically, the signal can be described by a function of the form\cite
{Demsar} $\psi (t,T)=A(t)\exp [i\omega _{A}(T)t+\phi ]+B\exp [t/\tau _{s}].$
The first term describes the AM oscillations, and the second term is the
contribution from single-particle relaxations. The amplitude of the AM decay
is well approximated by an exponential function $A(t)=A_{0}\exp [-t/\tau
_{p}]$ where $\tau _{p}$ is the AM decay time. We have found that by tuning
the laser wavelength near 800 nm, we can extinguish the second term almost
entirely, so that only the AM contribution is observed as shown in Figure
2a). The frequency $\omega _{A}(T)$ is temperature dependent, as shown in
Fig. 2b) which agrees well with previous Raman measurements\cite{coherent
phonons}. The discrete Fourier transform (FT) of $\psi $ in trace {\sf A} is
shown in Fig. 2c). Apart from the AM at $\omega _{A}=2.42$ THz, two
additional peaks are present, at 2.1 THz and 3.9 THz. These are attributed
to coherent phonon oscillations and are not of interest to us at present\cite
{coherent phonons}.

In Figure 2a) (trace {\sf B}) we also show the time-evolution of the AM\
after the application of a 2-pulse sequence (whose timing is indicated by
arows). The first pulse sets up the AM\ oscillatory state with $\phi =0$.
The second pulse is set to have equal amplitude, but is delayed with respect
to the first pulse such that its phase is shifted by $\pi $ with respect to
the first pulse, i.e. $\Delta t=\phi =[2n+1]\pi /\omega _{A}$. We see that
after the $\pi $ pulse in trace {\sf B }the amplitude of the AM\
oscillations is significantly reduced. From the FT of signal {\sf B} for $%
t>2 $ ps (shown in Figure 2c) it is evident that the amplitude of the AM is
reduced by more than 90\%, while the remaining signal contains significant
contributions of the coherent phonon oscillations at 2.1 and 3.9 THz, and of
course noise. From the point of view of data manipulation and computation,
we see that the AM has been effectively reset by the $\pi $ laser pulse.
Considering the fact that no spatial or temporal pulse shaping or amplitude
matching has been performed in these experiments, the efficiency of this
process is quite remarkable.

Similarly, we find that if the second pulse delay is set to be {\em in phase 
}with the first pulse\ $(\Delta t=\phi =2n\pi /\omega _{A}),$ the amplitude
is approximately doubled\ at $t=\phi $ by the second pulse, and the total AM
amplitude is $A=A_{1}+A_{2}$. This demonstration shows that $A$ and $\phi $
can be manipulated by the application of an appropriate laser pulse sequence
and allows any arbitrary superposition of states $\psi $ to be represented.

For the purpose of performing data processing operations, we can divide the
surface of the crystal into a set of $N=1...j$ discrete cells. Since the
frequency $\omega _{A}(T)$ of the AM is temperature dependent (Fig.2), we
can set the frequency of different cells with different $\omega _{j}$ by
adjusting the temperature of each cell. (This can be done with a temperature
gradient along the $x$ axis for example). This facilitates a tomographic
reconstruction of the database whereby different cells can be uniquely
identified by their frequency, and\ a Fourier transform (FT) of the readout
probe pulse which covers the entire array of cells simultaneously gives
directly the amplitude (and phase) of each cell. In effect it gives a
1-dimensional projection of $A$ and $\phi $ along the $x$ axis.

To implement a database search, we follow an operation sequence, which is
similar to a sequence given by Grover\cite{Grover2} invented for the case of
a quantum search. A database search operation is performed in two steps
preceeded by a setting up operation. First the amplitude $A_{j}\ $and phase $%
\phi _{j}$ of each cell is set up by an appropriately timed ''write'' laser
pulse ${\bf W.}$ This sets up the database $\Psi $ such that the information
in cell $j$ with frequency $\omega _{j}$ is represented by $A_{j}$ and $\phi
_{j}.$\ Since the time to perform the entire sequence of operations $%
t_{0}\ll \tau ,$ we assume that the amplitude is time independent $%
A_{j}(t)\simeq A_{j}$ and the reflectivity of each cell is thus given by$%
\;\psi _{j}=A_{j}\exp [i\omega _{j}t+\phi _{j}].$ The database is set up
such that one item is marked by setting its phase to be $\phi =\pi ,$ while
the unmarked cells have $\phi =0.$

The next step is to apply ${\bf \Pi }$ \ pulse on the entire array
simultaneously, which amplifies the set item and suppresses the rest. In our
example this is given by a properly timed laser pulse on the entire array,
such that $\Delta t=(2n+1)\pi /\omega _{0},$ where $\omega _{0}$ is the
center frequency of the array. The result is that the amplitude of the
marked cell is doubled, while the rest are suppressed. The reflectivity
modulations due to the entire array of cells is then read out by a pulse
beam $R,$ which covers the entire array. Thus the data encoded in the
surface of the crystal in cells $1...j$ is transferred to the reflected
readout beam whose intensity is proportional to $\Psi =\sum_{j}A_{j}\exp
[i\omega _{j}t+\phi _{j}]$ and contains all the phase and amplitude
information stored in the array, which can be displayed by a Fourier
transform of $\Psi .$

Here we limit\ ourselves to a two-element array $(N=2),$ which is sufficient
to effectively demonstrate the principle of operation. The first cell is
heated with a continuous wave argon laser, while the second is not, so that
the heated and unheated cells have frequencies of $\omega _{1}=2.38$ and $%
\omega _{2}=2.42$ THz respectively.

A write pulse ${\bf W}$ first excites the array, with $\phi _{1}=0$ for cell
1 and $\phi _{2}=\pi $ for cell 2$.$ The signal $\psi _{i}$ in the two cells
following the operation ${\bf W}\Psi $ is shown in Figure 3a). After $\Delta
t=2n\pi =2.09$ ps (corresponding to $n=5$) we apply a ${\bf \Pi }$ pulse.
The Fourier components ${\bf F}(\omega )$ at $\omega _{1}$ and $\omega _{2}$
before and after the $\Pi $ pulse are shown in Fig.3b). The amplitude is
obtained by a fit to the data to $\Psi =\psi _{1}+\psi _{2}=A_{1}\cos
[\omega _{1}t+\phi _{1}]+A_{2}\cos [\omega _{2}t+\phi _{2}]$\cite{FT}$.$
From the data shown in Figures 3a) and b) we see that the amplitude of $\psi
_{1}$ is approximately doubled, while the amplitude of $\psi _{2}$ is
extinguished, which, if applied to a larger array, would amount to
amplifying the set item and suppressing the rest. The identification of the
set item thus takes place in a single step. Clearly if $N>2$, application of
the same procedure would still yield the same result in a single step, as
long as there is only one set item\ in the database. The same search
algorithm is also applicable when more than one item is marked, and the
amount of information stored in each cell is limited by the resolution with
which the amplitude and phase can be read out. For a practically useful
system, to increase $N$ it would be more practical encode the data
2-dimensionally by using position rather than frequency to identify each
cell, enhancing the inherent parallelism of the processor. The crystal
surface is then divided into a 2D array, where the cells can be constructed
of arbitrary size, the lower limit being set by the coherence length of the
CDW\ system, $\xi \sim 10$ nm. With some adaptation, the optical elements
presently in use for temporal and spatial shaping of ultrashort optical
pulses \cite{Wefers2}, can then be used to control the timing of the
incident 2D array of pulses.

To conclude, the demonstrated ultrafast coherent-control data processor and
storage device using amplitude mode excitations of a CDW is still far from a
practical device. However, the parallel processing scheme using a CDW system
effectively demonstrates a conceptually new approch to ultrafast optical
data processing.

\bigskip

Figure 1 A schematic diagram of the experiments. The pulse timing is
controlled by variable time-delay units (Michaelson interferometers) TD$%
_{i}. $ M is a high-frequency acousto-optic modulator.

Figure 2 (a) The induced reflectivity $\psi =\Delta R(t)/R$ of the AM as a
function of time $t$ after excitation (trace {\sf A}). The figure shows the
raw data. Trace {\sf B} represents the reflecivity oscillations for a
2-pulse sequence. The timing of the second pulse is adjusted so that $\Delta
t=(2n+1)\pi /\omega _{A}$ (a $\pi $ pulse). (b) The $T$-dependence of $%
\omega _{A}$. (c) The discrete Fourier transforms of $\psi _{A}$ and $\psi
_{B}$ for $t>2.09$ ps shown in a). The component at $\omega _{A}$ is nearly
extinguished by the application of the $\pi $ pulse.

Figure 3 (a) a real-time display of the parallel search algorithm for a
2-cell array. ${\bf W}$ sets up the AM oscillation with $\phi _{1}=0$ in
cell 1 and $\phi _{2}=\pi $ in cell 2 at 2.38 and 2.42 THz respectively with
equal amplitudes. The operation ${\bf \Pi }$ amplifies the set data item (in
cell 2) while suppressing the amplitude in cell 1. The sum amplitude $\Psi $
is shown in the bottom trace.

(b) The Fourier amplitudes $F_{i}(\omega ),$ and phase $\phi _{i}$
(expressed both in terms of phase and time-delay) before and after the ${\bf %
\Pi }$ pulse are shown in the left and right panels respectively. After the $%
{\bf \Pi }$ pulse, the set item (in cell 2) is amplified, while the contents
(AM amplitude) of cell 1 are supressed.


\begin{references}
\bibitem{Demsar}  J.Demsar, K.Biljakovic and D.Mihailovic, Phys.Rev.Lett. 
{\bf 83}, 800 (1999)

\bibitem{Gruner}  See for example: G.Gr\"{u}ner, Density waves in solids,
(Addison Wesley, 1994) or G.Gr\"{u}ner, {\it Rev. Mod. Phys. }{\bf 60}, 1129
(1988)

\bibitem{Dekorsy}  T.Dekorsy, W.Kutt, T.Pfeifer, H.Kurz, Europhys. Lett. 
{\bf 23}, 223 (1993)

\bibitem{Dekorsy2}  A.Bartels, T.Dekorsy, H. Kurz, App. Phys. Lett. {\bf 72}%
, 2844 (1998)

\bibitem{Wefers}  M.M.Wefers, H.Kawashima, K.A.Nelso, J. Chem. Phys. {\bf 108%
}, 10248 (1998)

\bibitem{Hase}  M.Hase, K.Mizoguchi, H.Harima, S.Nakashima, M.Tani, K.Sakai,
M.Hangyo, App. Phys. Lett. {\bf 69}, 2474 (1996)

\bibitem{Falques}  J.Demsar and D.Mihailovic, in ''Spectroscopy of
superconducting materials'' Ed. E. Faulqes (American Chemical Society
Symposium series, 1999), p.230.

\bibitem{coherent phonons}  The frequencies of these modes correspond
closely to the observed in Raman spectroscopy. See for example S.Sugai,
Phys.Stat.sol. (b) {\bf 129}, 13 (1985).

\bibitem{Grover2}  L.K.Grover, Phys.Rev.Lett. {\bf 79}, 4709 (1997)

\bibitem{FT}  The same result could be obtained by performing a discrete
Fourier transform of the signals before and after the $\Pi $ pulse in Fig
3.98)

\bibitem{Wefers2}  M.M.Wefers, K.A.Nelso, A.M. Weiner, Opt. Lett. {\bf 21},
746 (1996)
\end{references}
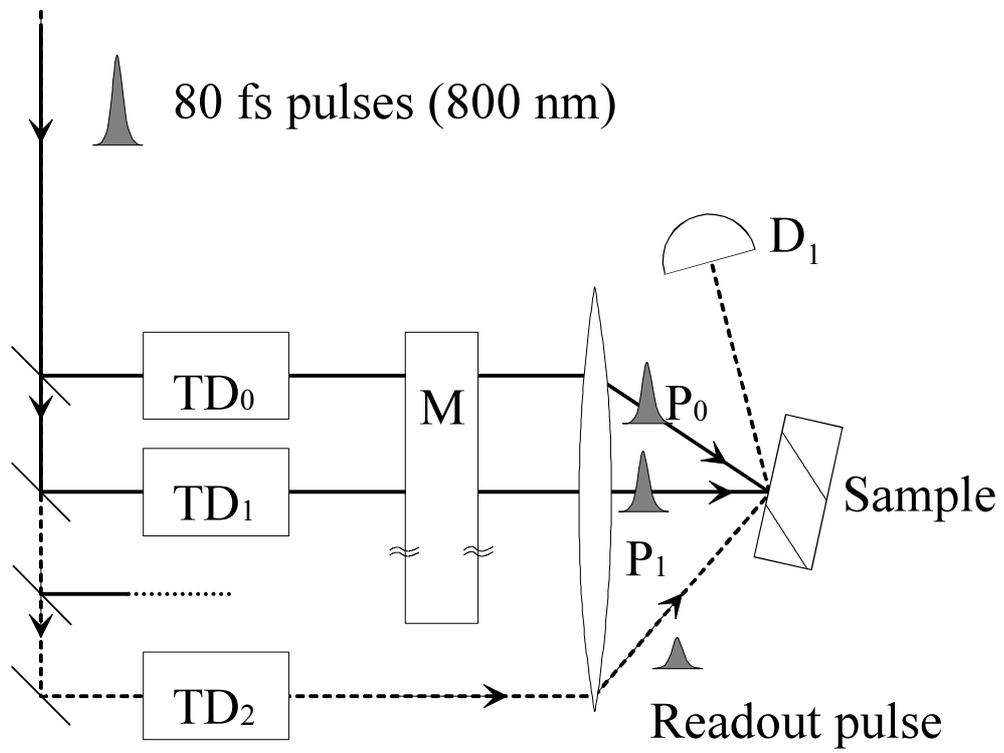
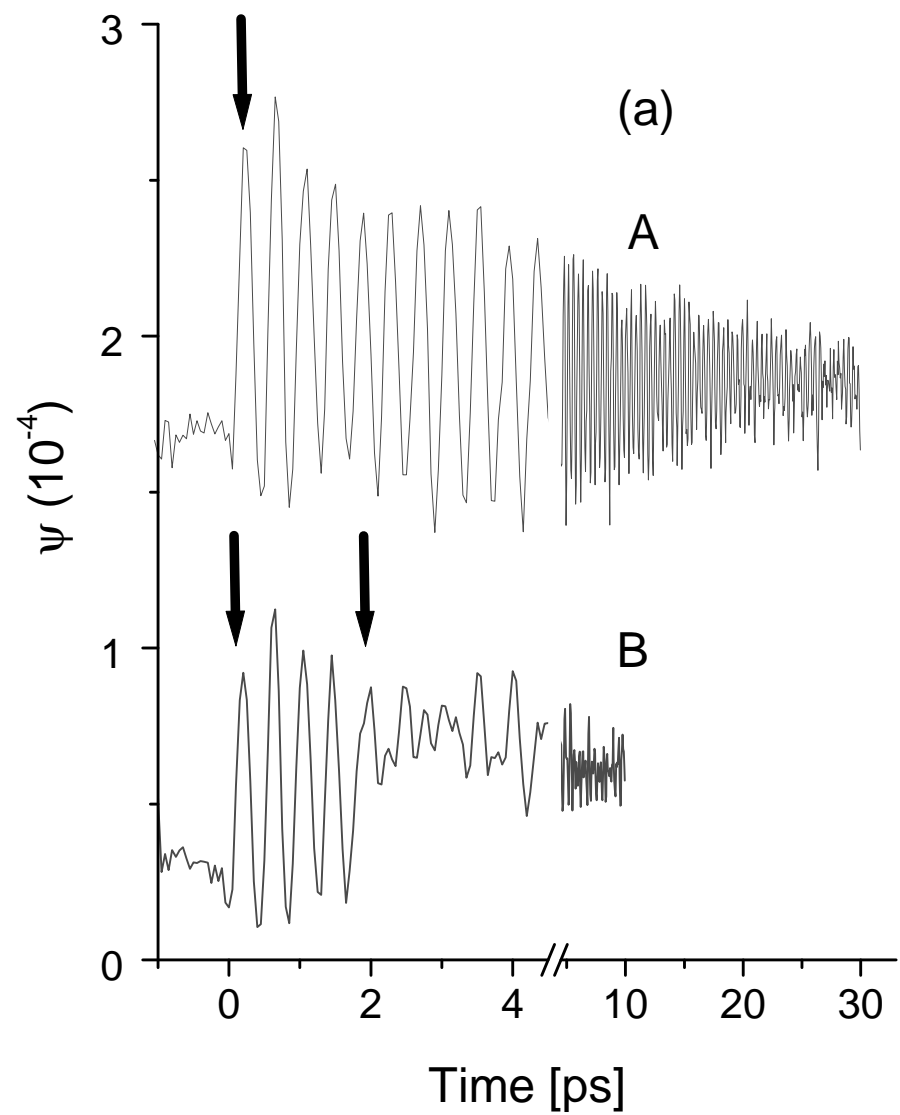
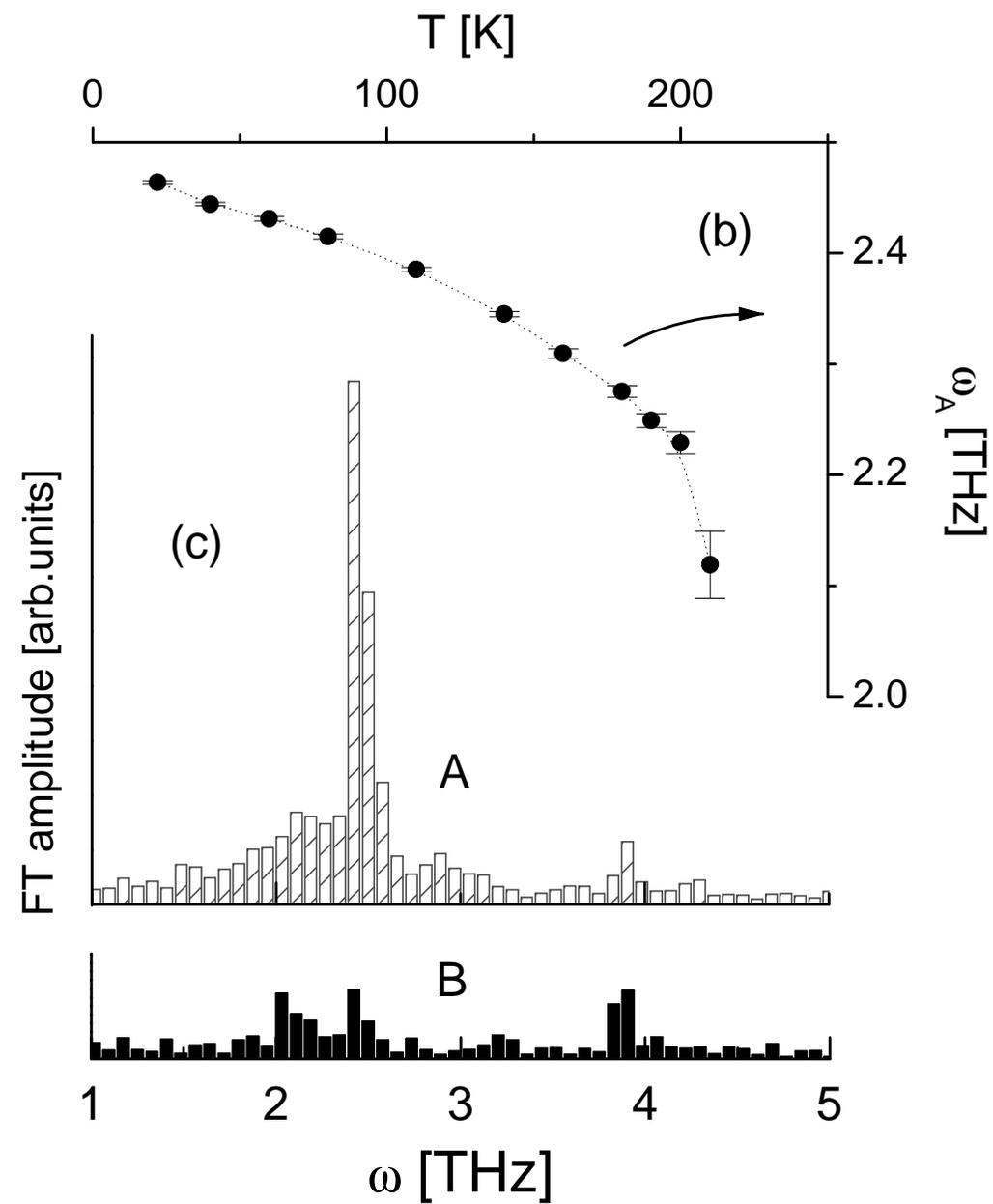
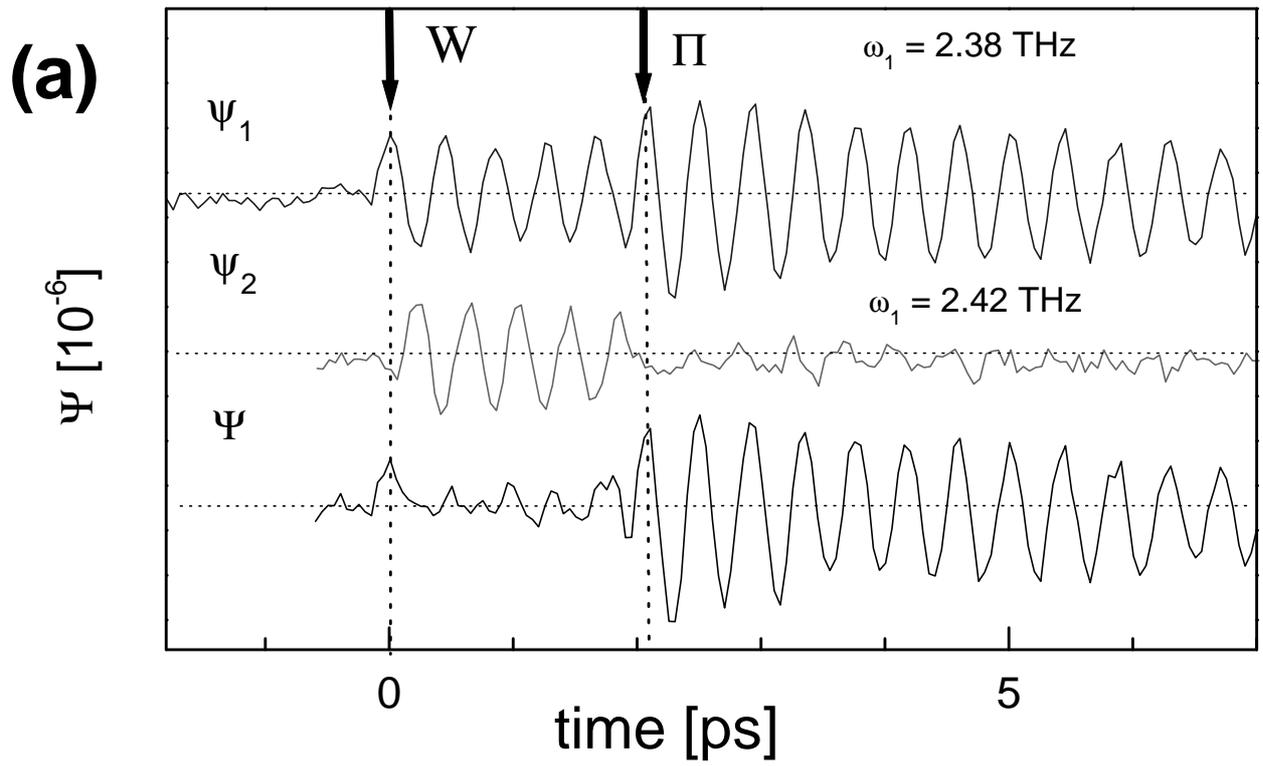
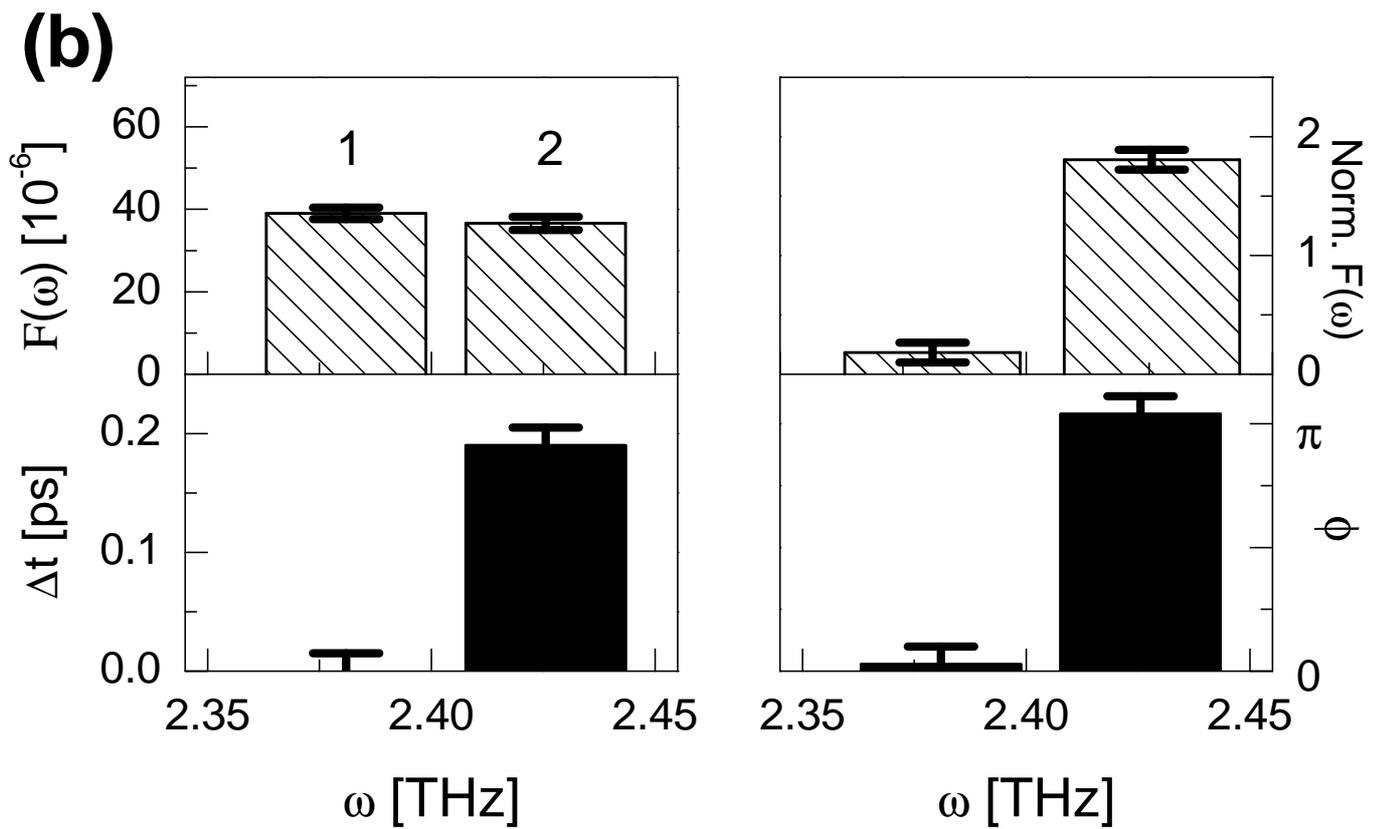
\end{document}